\documentclass[12pt, twoside]{article}
\pdfoutput=1
\usepackage{a4wide,amssymb,cite}
\usepackage{epsfig,graphicx}
\usepackage[usenames,dvipsnames]{color}
\usepackage{slashed}

\newcommand{\be}{\begin{equation}}
\newcommand{\ee}{\end{equation}}
\newcommand{\bea}{\begin{eqnarray}}
\newcommand{\eea}{\end{eqnarray}}

\begin{document}

\color{black}

\begin{flushright}
KIAS-P13026
\end{flushright}

\vspace{1cm}
\begin{center}
{\Large\bf\color{black}  Radiative generation of the Higgs potential}\\
\bigskip\color{black}\vspace{1.5cm}{
{\bf Eung Jin Chun, Sunghoon Jung, and Hyun Min Lee}
\vspace{0.5cm}
} \\[7mm]

{\it School of Physics, Korea Institute for Advanced Study, Seoul 130-722, Korea.  }\\
\end{center}
\bigskip
\centerline{\large\bf Abstract}
\begin{quote}
We consider the minimal extension of the Standard Model with a generalized $B-L$ gauge symmetry 
$U(1)_X$ for generating the Higgs potential radiatively.
Assuming that the full scalar potential vanishes at the vacuum instability scale, we achieve the goal in terms of two free parameters, the $X$ gauge coupling and the right-handed neutrino Yukawa coupling.
The $X$ gauge symmetry is broken spontaneously by the Coleman-Weinberg mechanism
while the scale symmetry breakdown induces electroweak symmetry breaking
through the radiative generation of appropriate scalar quartic couplings.
We show that there is a reasonable parameter space that is consistent with a correct electroweak symmetry breaking and the observed Higgs mass.

\end{quote}

\thispagestyle{empty}

\normalsize

\newpage

\setcounter{page}{1}

\section{Introduction}

The establishment of the Standard Model (SM), that requires a hypercharged doublet boson $H$ under the SM gauge symmetry $SU(2)_L\times U(1)_Y$ as the origin of the electroweak symmetry breaking (EWSB) and
all the quark and charged lepton masses, has culminated by the discovery of a new boson
at a mass around 126 GeV \cite{HiggsLHC}.
We now have fairly good information on the SM scalar potential describing the EWSB mechanism:
\be
V_H = m^2_H |H|^2 + \lambda_H |H|^4
\label{VH}
\ee
where $H$ can be written as $H=(0, v_{EW} + h)/\sqrt{2}$ in the unitary gauge.
The vacuum expectation value $v_{EW}$ and the Higgs particle mass $m_h$ are determined by
the two input parameters $m_H$ and $\lambda_H$ through the relation:
\be
m^2_H = - \lambda_H v_{EW}^2 = -{1\over2} M_h^2 \,,
\ee
which tells us $m_H^2 \approx - (89\, \rm{GeV})^2$ and $\lambda_H\approx 0.13$ translated from the
observed values of  $v_{EW} \simeq 246$ GeV and $M_h\approx 126$ GeV.

\medskip

The value of $\lambda_H\approx 0.13$
at the electroweak scale is quite intriguing as it is driven
to zero at a large scale close to the Planck scale \cite{degrassi12,3loops} and
develops the vacuum instability \cite{sher88}.
This may tell us about a possibility of generating
the Higgs quartic coupling radiatively starting from the vanishing initial condition at a high scale \cite{shapos09}.
The electroweak scale $v_{EW}$ (or $m^2_H$) might also have a radiative origin
as was shown decades ago by Coleman and Weinberg \cite{CW} that a mass scale
can be generated via dimensional transmutation from running couplings in a massless scalar
gauge theory.  Combining these two features, one may envisage a possibility that
the whole Higgs potential (\ref{VH}) is dynamically generated.

\medskip

While the Coleman-Weinberg (CW) mechanism does not work for the SM,
it is conceivable to extend the SM with an additional scalar $U(1)$ gauge theory
in which the additional $U(1)$ symmetry is spontaneously broken by radiative corrections
and the generated mass scale is transferred to the SM  \cite{hemp96,nicolai,wu,iso09,iso12,englert13}.
Thus, in this paper, we consider the possibility of implementing the CW mechanism
to a generalized $B-L$ gauge symmetry, $U(1)_X$, which is a natural extension of the SM with
three right-handed neutrinos to explain the observed neutrino masses and mixing. 
Here, the charge  operator $X$ is taken as a linear combination of the conventional $B-L$ charge operator $Y_{B-L}$ and the hypercharge operator $Y$:
\begin{equation}
 X = Y_{B-L} - x Y
\end{equation}
where $x$ is a real number.  We can freely choose a charge mixing parameter $x$ within a range satisfying a certain criterion that will be discussed later.

The $B-L$ extended SM contains an additional complex scalar field $\Phi$ whose vacuum expectation value (VEV) breaks the $U(1)_X$ gauge symmetry 
and induces Majorana masses of right-handed neutrinos
$M_N = y_N \langle \Phi \rangle$ through the Yukawa coupling $y_N$.
Once a non-vanishing $\langle \Phi\rangle$ is generated, it can generate the Higgs mass by
$m^2_H= \lambda_{H\Phi} \langle \Phi\rangle^2$ in the presence of
the mixing potential term $\lambda_{H\Phi} |\Phi|^2 |H|^2$.
As will be seen,
it is remarkable that the coupling $\lambda_\Phi$ of the quartic potential term
$\lambda_\Phi |\Phi|^4$ is radiatively generated by the right-handed neutrino Yukawa coupling $y_N$ in a similar way as in the Higgs quartic coupling in the SM. But, there is a difference that the beta function of $\lambda_\Phi$ changes sign during the renormalization group (RG) evolution, unlike the monotonic behavior of the running Higgs quartic coupling $\lambda_H$ due to the top Yukawa coupling.
That is, $\lambda_\Phi$ is generated
radiatively below the instability scale dominantly by a sizable right-handed neutrino Yukawa coupling but it becomes small enough for the Coleman-Weinberg mechanism to work at even lower scales.


\medskip

Putting together all the features discussed above, we wish to entertain a paradigm of radiative generation
of all the scalar potential terms which are supposed to vanish at a certain UV scale.
To be specific, we  consider the possibility of achieving a spontaneous breaking of the electroweak and generalized $B-L$ gauge symmetries by the addition of only two free parameters, the right-handed neutrino Yukawa coupling $y_N$ and the $U(1)_X$ gauge coupling $g_X$.

The paper is organized as follows.
We begin with the description of the $B-L$ extension of the SM and review the Coleman-Weinberg potential in the model.
Then we perform the RG analysis of the model and search the parameter space that give rises to a correct electroweak symmetry breaking and Higgs mass.
Finally, conclusions are drawn. There is one appendix containing the RG equations of the model.

\section{General $B-L$ extension of the SM and Coleman-Weinberg potential }

The $B-L$ extension of the Standard Model that we are considering is described by the following Lagrangian \cite{B-L},
\be
{\cal L}={\cal L}_S +{\cal L}_{\rm YM}+ {\cal L}_F +{\cal L}_Y,
\ee
with
\bea
{\cal L}_S&=& |D_\mu H|^2 +|D_\mu \Phi|^2 - m^2_H |H|^2 - m^2_\Phi |\Phi|^2 -\lambda_H |H|^4 -\lambda_\Phi |\Phi|^4 -\lambda_{H\Phi} |H|^2 |\Phi|^2, \\
{\cal L}_{ \rm YM}&=& -\frac{1}{4} (F^{\mu\nu} F_{\mu\nu})_{\rm SM} -\frac{1}{4} F^{\prime \mu\nu} F'_{\mu\nu}, \\
{\cal L}_F &=& i {\bar q}_L \slashed{D} q_L+i{\bar u}_R \slashed{D} u_R +i {\bar d}_R \slashed{D} d_R + i {\bar l}_L \slashed{D} l_L+i {\bar e}_R \slashed{D} e_R + i {\bar \nu}_R \slashed{D} \nu_R, \\
{\cal L}_Y &=& - y_d {\bar q}_L d_R H-y_u {\bar q}_L u_R {\tilde H}-y_e {\bar l}_L e_R H -y_\nu {\bar l}_L \nu_R {\tilde H} - y_N \overline {(\nu_R)^c} \nu_R \Phi   \label{yukawas}
\eea
where ${\tilde H}=i\sigma^2 H^*$ and the covariant derivative is
\be
D_\mu = \partial_\mu + ig_S T^\alpha G^\alpha_\mu + ig T^a W^a_\mu +i g_Y Y B_\mu + i({\tilde g} Y+g_X X) B'_\mu.
\ee
Note that the gauge coupling $\tilde g$ describes the kinetic mixing between $U(1)_Y$ and $U(1)_X$.
In Table 1, we show $Y$ and $Y_{B-L}$ charges
as well as $X$ charges for a representative $U(1)_X$ with $x=4/5$ which will be used below
to show how the radiative generation of the full scalar potential works.  When the complex scalar $\Phi$, carrying the $X$ number 2, gets a VEV, the $U(1)_X$ gauge symmetry is broken spontaneously and the right-handed neutrinos obtain masses.

\begin{table}
\begin{center}
\begin{tabular}{|c|c|c|c|c|c|c|c|c|}
\hline
 &  $q_L$ & $u_R$  & $d_R$ & $l_L$  & $e_R$ & $\nu_R$ & $H$ & $\Phi$ \\
\hline
$Y$ & ${1\over6}$ & ${2\over3}$ & ${-1\over3}$ & $-{1\over2}$  & $-1$ & $0$ & ${1\over2}$ & $0$ \\
\hline
$Y_{B-L}$ & ${1\over3}$ & ${1\over3}$ & ${1\over3}$ & $-1$  & $-1$ & $-1$ & $0$ & $2$ \\
\hline
$X_{4\over5}$ & ${1\over5}$ & $-{1\over5}$ & ${3\over5}$ & $-{3\over5}$ & $-{1\over5}$ & $-1$ & $-{2\over5}$ & $2$ \\
\hline
\end{tabular}
\caption{Quantum numbers of particles in the Standard Model extended 
with a generalized $B-L$ gauge symmetry $U(1)_X$ with $x=4/5$.  }
\end{center}
\end{table}

\medskip

Now we consider the one-loop Coleman-Weinberg potential \cite{CW} for the $B-L$ sector.
In the limit of $|\lambda_{H\Phi}|, |{\tilde g}|  \ll 1$ at tree level, we can ignore the SM contributions to the CW potential and focus on the $U(1)_X$ sector.
Taking $\Phi=\phi/\sqrt{2}$ in the unitary gauge and
imposing the renormalization conditions \cite{CW},
\bea
\frac{\partial^2 V}{\partial \phi^2}\Big|_{\phi=0}&=&0, \\
\frac{\partial^4 V}{\partial \phi^4}\Big|_{\phi=M}&=& 6\lambda_\Phi,
\eea
the one-loop corrected $U(1)_X$ potential becomes\footnote{The factor in the $y^4_N$ term is corrected as pointed out in Ref.~\cite{hashimoto13}.}
\be
V_X(\phi)=\frac{1}{4}\lambda_\Phi \phi^4 +\frac{\phi^4}{64\pi^4}\Big(10\lambda^2_\Phi+48 g^4_{B-L} -8\sum_{i=1}^3y^4_{N_i}\Big)\bigg(\ln \frac{\phi^2}{M^2}-\frac{25}{6}\bigg)
\label{CWpot}
\ee
where we assumed that the Yukawa couplings for the right-handed neutrinos are diagonal as $y_{N,ij}=y_{N_i} \delta_{ij}$. For our analysis in the following, we will take only one coupling $y_N$.

Choosing the renormalization scale at
$M=\langle\phi\rangle\equiv v_\phi$ to avoid the large-log uncertainty in the one-loop approximation \cite{CW}, one can evaluate the minimization condition of the potential  (\ref{CWpot}) and obtain
\be
\lambda_\Phi(v_\phi)=\frac{11}{48\pi^2}\Big(10\lambda^2_\Phi+48 g^4_X - 
8 y^4_{N}\Big)(v_\phi).
\label{Vmin}
\ee
This relation fixes the $U(1)_X$ breaking scale $v_\phi$ in terms of input values of
$\lambda_\Phi$, $g_X$ and $y_N$ which evolve from a high scale $M_*$ to $v_\phi$ by RG.
As a consequence, the CW potential (\ref{CWpot}) leads to a naturally small VEV $v_\phi$ via dimensional transmutation as
\be
v_\phi \simeq M_*\,e^{\frac{11}{6}}\, {\rm exp}\bigg(-\frac{\pi^2}{6}\frac{\lambda_\Phi(M_*)}{g^4_{B-L}(M_*)-\frac{16}{96}y^4_{N}(M_*)}\bigg)  \label{dimtrans}
\ee
where the small $\lambda_\Phi^2$ contribution is neglected and $g^4_X-\frac{1}{6}y^4_{N}>0$.
When the beta function of $\lambda_\Phi$ changes the sign during the RG evolution, we should take $M_*$ to be below the scale where $g^4_X-\frac{1}{6}y^4_{N}=0$.  As will be seen in the later section, even if we start with a vanishing $\lambda_\Phi$ at the cutoff scale, a positive $\lambda_\Phi$ is generated at a smaller scale by the RG evolution with a positive beta function of $\lambda_\Phi$, setting the initial couplings for dimensional transmutation given in eq.~(\ref{dimtrans}).

\medskip

We also obtain the physical masses of the $U(1)_X$ scalar $\phi$ and the gauge boson $B'$
in the vacuum as
\bea
M^2_\phi = \frac{6}{11}\lambda_\Phi(v_\phi)v^2_\phi\quad\mbox{and}\quad
M^2_X=4 g^2_X(v_\phi) v^2_\phi\,,
\label{singletmass}
\eea
which determines the ratio $M_\phi^2/M^2_X \approx 3 (1-y_N^4/6 g^4_X) g^2_X/2 \pi^2$ putting a rough relation (\ref{Vmin}),
and thus $M_\phi \ll M_X$.


\section{Radiative $B-L$ and electroweak symmetry breaking}

The discovered Higgs boson has a mass at $126\,{\rm GeV}$, so the Higgs quartic coupling vanishes at $M_I$, the so called vacuum instability scale\footnote{The vacuum instability scale is essentially regarded as the UV cutoff but we don't specify a UV completion for curing the vacuum instabilty. It could be a field-theoretical UV completion with heavy particle or a quantum gravity. The former case could reintroduce the hierarchy problem via the Higgs coupling.}, which is below the Planck scale, for the top pole mass $M_t>171\,{\rm GeV}$ \cite{degrassi12,3loops}.
If there is a sizable new physics contribution to the running of the Higgs quartic coupling, it is possible to increase the vacuum instability scale. But, in order not to reintroduce the hierarchy problem, a new particle mass curing the vacuum instability should not be far away from the weak scale \cite{Sthreshold}. In our case, we assume that there is no heavy particle other than the $B-L$ sector. Then, the vacuum instability scale remains the same as in the SM.

\medskip

Generalizing the vanishing feature of the Higgs quartic coupling at $M_I$,
we assume the initial condition of vanishing all the scalar potential terms as well as the kinetic mixing coupling $\tilde{g}$ in a certain UV completed theory\footnote{ We note that the two initial conditions of $\tilde g=0$ and vanishing scalar potential are characteristically different. The first would arise, for instance, from breaking a simple unified gauge group $G_{\rm GUT}$ into $SU(3)_C\times SU(2)_L\times U(1)_Y\times U(1)_X$. }
 at $M_I$,
 and examine whether it leads to viable electroweak and $U(1)_X$ symmetry breaking by solving the RG equations presented in the Appendix.
That is, we impose the initial condition at $M_I$:
\be
\lambda_H=0,\; \lambda_{\Phi}=0,\; \lambda_{H\Phi}=0,\;
m^2_\Phi=0, \; m^2_H=0, \;\mbox{and}\; {\tilde g}=0,
\ee
with arbitrary values of $y_N$ and $g_X$, and then calculate the RG generated values at lower scale
to determine the $U(1)_X$ breaking scale $v_\phi$ at which the relation (\ref{Vmin}) is satisfied.
During the RG evolution, a small negative $\lambda_{H\Phi}$ is generated at one-loops
involving $(\tilde{g}-x g_X)^2 g^2_X$ and thus the Higgs potential develops at $v_\phi$ as
\be
V_{\rm SM}(h)= \frac{1}{2}m^2_H(v_\phi) h^2+\frac{1}{4} \lambda_H(v_\phi) h^4
\ee
with
\be
m^2_H(v_\phi)= \frac{1}{2}\lambda_{H\Phi}(v_\phi) v^2_\phi \,.
\ee
Below $v_\phi$, the Higgs quartic coupling $\lambda_H$ runs approximately as in the SM
while the running of the Higgs mass parameter is governed by
\be
\frac{d m^2_H}{d\ln\mu}= \frac{1}{16\pi^2}\bigg[m^2_H\Big(12\lambda_H+6 y^2_t -\frac{9}{2}g^2-\frac{3}{2}g^2_Y -\frac{3}{2}({\tilde g}-x g_X)^2 \Big)+2\lambda_{H\Phi} M^2_\phi\bigg]
\label{mHrun}
\ee
where $M^2_\phi$ is given in eq.~(\ref{singletmass}).
Note that the right-hand side of (\ref{mHrun}) contains an additive term proportional to a new scalar mass-squared $M^2_\phi$ given by (\ref{singletmass}). But it gives a contribution
of order $\lambda_\Phi m^2_H$ which is negligible for $\lambda_\Phi\ll 1$.
Then, the electroweak VEV is determined by
\be
v_{EW}= \sqrt{-\frac{m^2_H(v_{EW})}{\lambda_H(v_{EW})}}.
\ee
On the other hand, the Higgs mass is given by
\be
M^2_h= 2\lambda_H(v_{\rm EW}) v^2_{EW}+\Delta M^2_h
\ee
where $\Delta M^2_h$ is the Higgs self-energy correction to the Higgs pole mass \cite{degrassi12}.
Thus, one can select out appropriate initial values of $y_N$ and $g_X$ reproducing
two observables, $v_{EW}\simeq 246$ GeV and $M_h\approx 126$ GeV.
We consider two UV scales, $M_I = 2\times 10^{11}$ and $10^{18}$ GeV, corresponding to the vacuum
instability scales for the top mass $M_t=173$ and $171.3$ GeV, respectively.

\begin{figure}[t]
\centering
\includegraphics[width=0.49\textwidth]{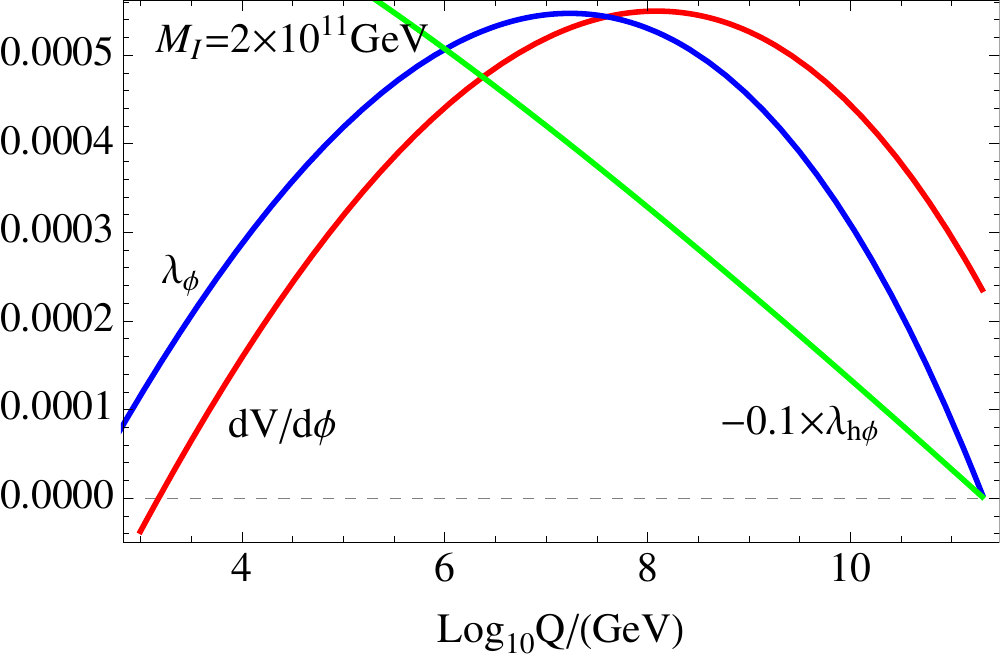}
\includegraphics[width=0.49\textwidth]{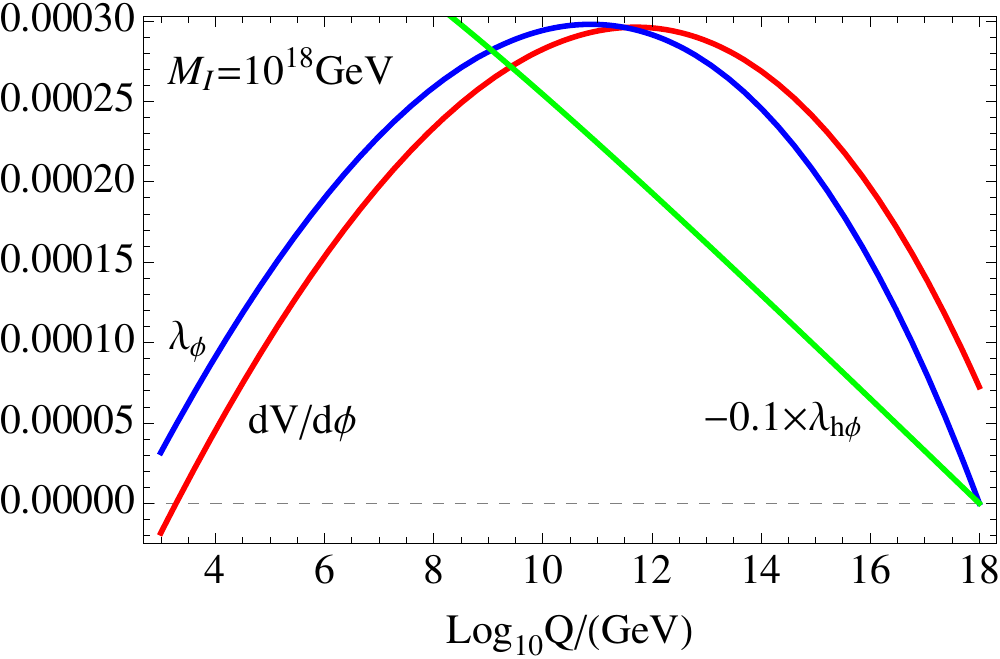}
\caption{Examples of running  quartic coupling $\lambda_\Phi$ (Blue),  the minimization condition (\ref{Vmin}) (Red), and $\lambda_{H\Phi}$(Green) (multiplied by $-0.1$ to fit in the plot) for the instability scale $M_I=2\times10^{11}\,{\rm GeV}$ on the left and $M_I=10^{18}\,{\rm GeV}$ on the right.
Successful electroweak symmetry breaking occurs in both examples with the charge mixing parameter, $x=4/5$.
}
\label{fig:lphirun}
\end{figure}

\begin{figure}[t]
\centering
\includegraphics[width=\textwidth]{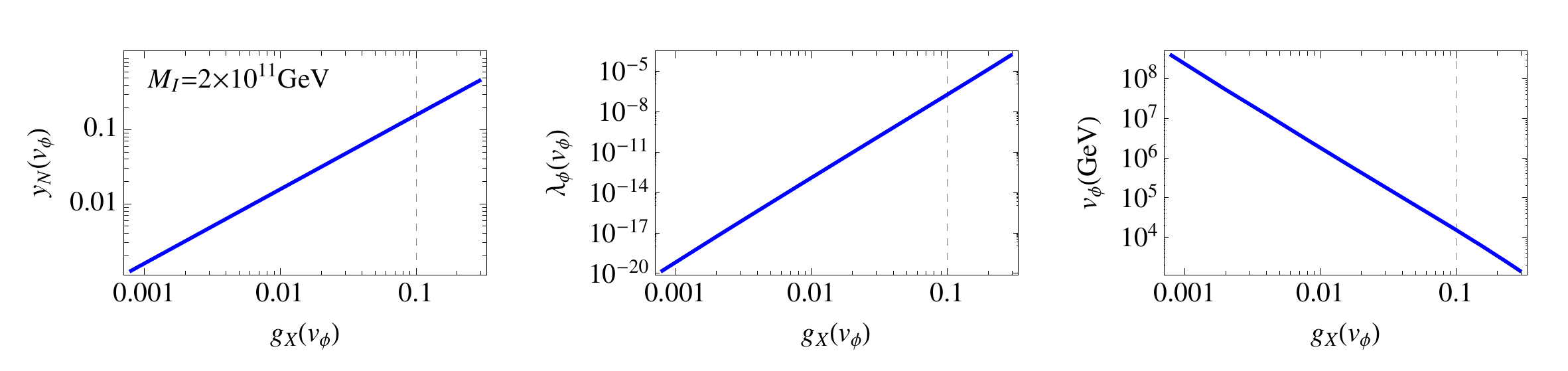}
\includegraphics[width=\textwidth]{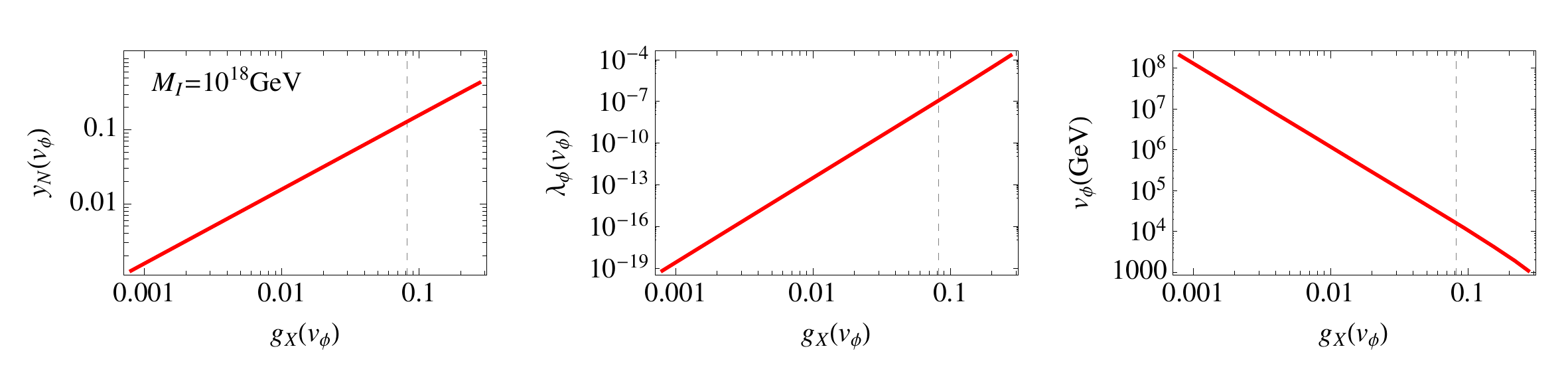}
\caption{The values of the gauge coupling $g_{X}$ {\it vs.}\ the right-handed neutrino Yukawa coupling $y_N$ (Left), the quartic coupling $\lambda_\Phi$ (Middle), and the $U(1)_X$ breaking scale $v_\phi$ (Right) realizing successful electroweak symmetry breaking. We have chosen the charge mixing parameter to $x=4/5$, the Higgs mass at $126\,{\rm GeV}$ and the instability scale, $M_I=2\times 10^{11}\,{\rm GeV}$ and $10^{18}\,{\rm GeV}$, in the upper and lower panels, respectively. We get $M_X>3$TeV in the region left to the vertical dashed line.
}
\label{fig:parameters}
\end{figure}
\begin{figure}[t]
\centering
\includegraphics[width=\textwidth]{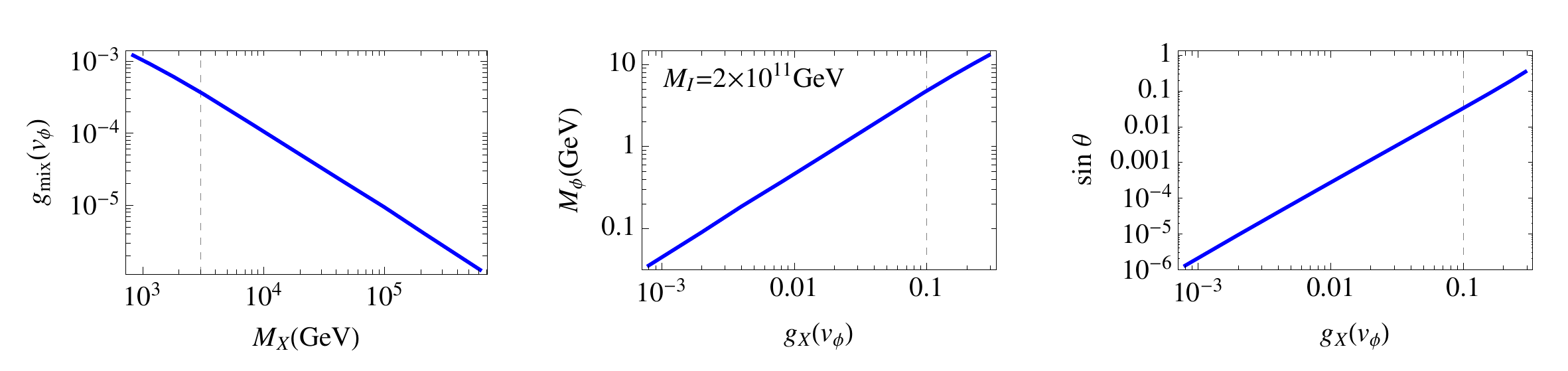}
\includegraphics[width=\textwidth]{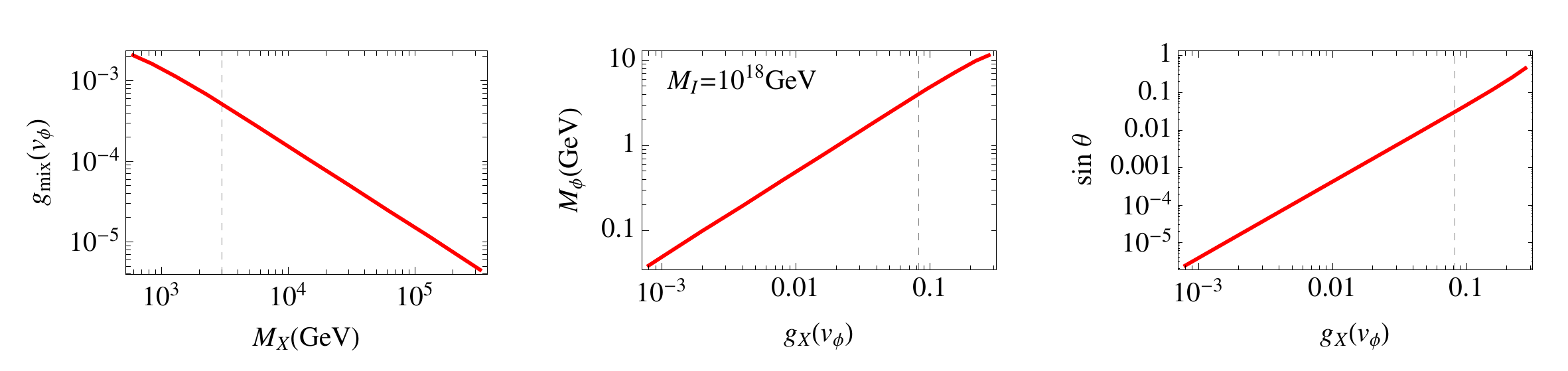}
\caption{
The values of the $U(1)_X$ gauge boson mass {\it vs.}\ the kinetic mixing, $g_{\rm mix}={\widetilde g}$ (Left).
The values of the gauge coupling $g_X$ {\it vs.}\ singlet scalar mass $M_\phi$ (Middle) and Higgs mixing parameter $\sin\theta$ (Right). We have chosen the charge mixing parameter to $x=4/5$, the Higgs mass at $126\,{\rm GeV}$ and the instability scale, $M_I=2\times 10^{11}\,{\rm GeV}$ and $10^{18}\,{\rm GeV}$, in the upper and lower panels, respectively. The vertical dashed line corresponds to $M_X=3\,{\rm TeV}$.
}
\label{fig:singlet}
\end{figure}

\medskip

Figure 1 shows examples of radiative generation of the $U(1)_X$ quartic coupling 
$\lambda_\Phi$ and
the $U(1)_X$ symmetry breaking scale for appropriate choices of $y_N$ and $g_X$ at
$M_I = 2\times 10^{11}$ and $10^{18}$ GeV, respectively. 
Also shown is the running of the mixing quartic coupling between $H$ and $\Phi$, which is generated mainly by the $g_X^4$
term for $x\neq 0$ as can be seen from (A.10).
 In both examples, the beta function of $\lambda_\Phi$ starts with a negative sign at the instability scale so a positive $\lambda_\Phi$ is generated at a smaller scale, setting the initial parameters for a dimensional transmutation via eq.~(\ref{dimtrans}).
But, the beta function of $\lambda_\Phi$ at even smaller scales becomes positive, resulting in a small $\lambda_\Phi$ appropriate for satisfying the $U(1)_X$ minimization condition, $V'(\Phi)=0$.

 In our scheme, a general criterion for successful CW minimization has been derived in Ref.~\cite{hashimoto13}. For the generalized $B-L$ gauge symmetry $U(1)_X$ with arbitrary $x$, we have $K=(108 -64 x +41 x^2)/36\sqrt{6}$ which becomes less than 1 for $0.43<x<1.13$ to allow the dynamical breaking of $U(1)_X$ symmetry with a vanishing potential in the UV.
 For our analysis, a specific value of $x=4/5$ is taken [see Table 1] to show how the whole mechanism works. 

\medskip

The results of our analysis are summarized in Figure 2 and 3. 
Figure 2 shows the lines in the plane of
the $U(1)_X$ gauge coupling $g_X$ and the right-handed neutrino Yukawa coupling $y_N$ determined
at the $U(1)_X$ symmetry breaking scale $v_\phi$, which satisfy the electroweak symmetry breaking conditions.  Also shown are the induced values of $\lambda_\Phi$ and $v_\phi$ as functions of $g_X$.
The upper (lower) panels are for $M_I=2\times10^{11}$  ($10^{18}$) GeV.
From these plots, one can read off the predicted mass scales: $M_X = 2 g_X v_\phi$ for the $U(1)_X$ gauge boson, $M_\phi = \sqrt{6\lambda_\Phi/11} v_\phi$ for the $U(1)_X$ scalar, and $M_N= \sqrt{2} y_N v_\phi$ for the right-handed neutrino.
The $U(1)_X$ breaking scale lies in the range $10^4\,{\rm GeV}\lesssim v_\phi \lesssim10^8\,{\rm GeV}$ for $10^{-3}\lesssim g_X\lesssim 0.1$ and it gets smaller for higher $M_I$. We note that $v_\phi$ as low as a few TeV can be obtained for $g_X(v_\phi)\gtrsim  0.2$ for which
the $U(1)_X$ gauge boson signatures may be found in the future LHC run.
One can see from Figure 2 that the radiative breaking of the $U(1)_X$ and electroweak symmetries occurs appropriately in a reasonable range of the two input parameters $y_N$ and $g_X$.
However, one finds that $\lambda_\Phi \ll g_X^4$ which requires a fine cancellation between $g_X^4$ and $y_N^4$ contributions in the minimization condition. We note that from (A.8), the beta function of the Higgs quartic coupling acquires an additional contribution proportional to $x^4 g^4_X$ as compared to the standard $B-L$ symmetry. But, since the gauge coupling $g_X$ is rather small, $g_X\lesssim 0.1$, for $U(1)_X$ symmetry breaking, the running of the Higgs quartic coupling is almost the same as in the SM.

In Figure 3, we show the values of  the kinetic mixing $\tilde g=g_{\rm mix}$ as a function of the $U(1)_X$ gauge boson mass $M_X$, and the values of the physical singlet scalar mass $M_\phi$ and the Higgs mixing angle $\sin\theta$ as a function of the gauge coupling $g_X$. All the values satisfy the electroweak symmetry breaking conditions. Note that the small kinetic mixing ($\tilde g \ll g_X$) plays an unimportant role in the $U(1)_X$ scheme with $x \sim 1$ as its contribution to the running of $\lambda_{H\Phi}$ is subdominant to that of $g_X$.
In the case of $M_I=2\times 10^{11}\,{\rm GeV}$,  the singlet scalar mass ranges  between $0.1-8\,{\rm GeV}$ and the Higgs mixing is about $\sin\theta\sim 5\times 10^{-4}-10^{-3}$ for $g_X (v_\phi) =0.002-0.2$. On the other hand, in the case of $M_I=10^{18}\,{\rm GeV}$, we obtain even lighter singlet scalar masses and relatively larger mixing. Thus, smaller $v_\phi$ successfully triggers electroweak symmetry breaking.

Note that the resulting $\lambda_\Phi$ and $M_\phi$ become much smaller than in the case with $x=0$.  For $g_X(v_\phi)\sim 0.1$, $M_X \approx 0.9 M_N$ can be multi-TeV while $M_\phi$ is only a few GeV. Such a light scalar can be produced by the Standard Model Higgs boson
decay $h \to \phi \phi$ through small mixing $\sin\theta$ although being too small to observably affect Higgs decays, or by the $X$ boson decay $X\to \phi\phi$.
As $\phi$ decays mainly to $\tau\bar\tau$ or $c\bar c$, observing a very narrow resonance in these final states would provide an interesting signal of the radiative generation mechanism of the Higgs potential

\medskip

Let us finally comment on the possibility of allowing the right-handed neutrino mass higher than $10^9$ GeV for which successful thermal leptogenesis can easily occur \cite{DI}.
In order to achieve the EWSB with a high $B-L$ breaking scale $v_\phi>10^9 \,{\rm GeV}/(\sqrt{2}y_N)$, one needs to generate an extremely small mixing coupling $|\lambda_{H\Phi}| < 10^{-14}$. This requires an unnaturally small value of  $g_X \lesssim 3\times10^{-4}$ as the RG induces roughly $|\lambda_{H\Phi}| \sim g_X^4 \ln(M_I/v_\phi)/\pi^2$.
Furthermore, we also note that the neutrino Yukawa coupling $y_\nu$ to the SM Higgs in eq.~(\ref{yukawas}) leads to an additional term in the beta function of the Higgs mass parameter (\ref{mHrun}), $\Delta\beta_{m^2_H}=8\lambda^2_N M^2_N/(16\pi^2)$.
Thus, in order for the extra Higgs mass term not to exceed the one from the mixing Higgs quartic coupling, one needs $M_N\lesssim |m_H|(16\pi^2 |m_H|/m_\nu)^{1/3}$ where use is made of the see-saw formula, $m_\nu=y^2_\nu v^2/M_N$, and gets the upper bound on
the $U(1)_X$ breaking scale, $v_\phi \lesssim10^7\,{\rm GeV}/(\sqrt{2}y_N)$, for $m_\nu\simeq 0.05\,{\rm eV}$ \cite{iso09,finetune}.
Therefore, the standard thermal leptogenesis could be in tension with
the Higgs mass bound. However, a marginal compatibility can be found if right-handed neutrino domination is assumed at high temperature in which case a successful leptogenesis occurs for $M_N>2 \times10^7\,{\rm GeV}$ \cite{lowerMN}. Note that there is a two-loop contribution to the Higgs mass parameter due to top and $U(1)_X$ gauge boson: $\Delta m^2_H\sim y^2_t \alpha^2_X M^2_X/(16\pi^2)$ \cite{iso09}. Again, from $\Delta m^2_H\lesssim m^2_H$, we obtain the upper bound on the $U(1)_X$ breaking scale as $v_\phi\lesssim (0.1/g_X)^3\times 10^7\,{\rm GeV}$ which is well above
our solution line in Figure 2.

\section{Conclusion}

The origin of the EWSB would be a fundamental question in the SM, which may be related to new physics explaining
neutrino masses and/or dark matter. An appealing way of generating a mass scale dynamically is the Coleman-Weinberg mechanism which is known to work in a massless scalar $U(1)$ gauge theory.
It is then tempting to consider a $U(1)$ symmetry in connection with neutrino mass generation and/or
the existence of dark matter.  As one of the examples, we considered the radiative breaking of the generalized $B-L$ symmetry $U(1)_X$,  whose generator $X$ is a linear combination of the conventional $B-L$ charge and the hypercharge $Y$, 
which can lead to right-handed neutrino masses and the Higgs mass at the same time. We note that the $U(1)_X$ scalar
quartic coupling, although vanishing at a high scale, can be generated dynamically due to the right-handed neutrino Yukawa coupling $y_N$,
just as the SM scalar quartic coupling is generated due to the top quark Yukawa coupling below the instability scale.

Performing the RG analysis in the $U(1)_X$ extended SM, we examined the possibility of generating radiatively
the whole scalar potential at low scales for the initial condition of a vanishing potential at the instability scale.
A small quartic coupling $\lambda_\Phi $ required for the Coleman-Weinberg generation of the $U(1)_X$ scalar VEV is obtained even for a vanishing initial $\lambda_\Phi$ due to the sign change of the corresponding beta function.
We also found that a naturally small mixing coupling of the $U(1)_X$ and electroweak Higgs 
scalars is generated through small $U(1)_X$ gauge coupling $g_X$.
We showed that there are reasonable values of $y_N$ and $g_X$ for the successful $U(1)_X$ and electroweak symmetry breaking, being consistent with the measured Higgs boson mass.

\section*{Acknowledgments}
We would like to thank Dumitru Ghilencea for the discussion related to a scale-invariant Standard Model.
EJC was supported by the National Research Foundation of Korea (NRF) grant funded by the Korea government (MEST) (No.~20120001177).
HML thanks Theory Division at CERN for a warm hospitality during his visit.

\def\theequation{A.\arabic{equation}}

\setcounter{equation}{0}

\vskip0.8cm
\noindent
{\Large \bf Appendix A:  Renormalization group equations}
\vskip0.4cm
\noindent

The running of couplings $p_i$ are governed by the RG equations, $\frac{d p_i}{dt}=\beta_{p_i}$ with $\beta_{p_i}$ being the corresponding beta functions  and $t\equiv\ln(Q/M_t)$. We present the beta functions in the $B-L$ extension of the SM \cite{B-L} with the correction of the factor 16 
in the $y^4_N$ term for $\beta_{\lambda_\Phi}$ \cite{hashimoto13}.

First, the one-loop beta functions of the gauge couplings are
\bea
(4\pi)^2 \beta_{g_Y}&=&\frac{41}{6} g^3_Y, \quad (4\pi)^2 \beta_{g}= -\frac{19}{6} g^3,  \quad
(4\pi)^2 \beta_{g_S}=-7 g^3_S, \\
(4\pi)^2 \beta_{g_X}&=& \Big(12 -\frac{32}{3}x+\frac{41}{6}x^2\Big) g^3_X
+\Big(\frac{32}{3}-\frac{41}{3}x \Big)g^2_X\, {\widetilde g}+\frac{41}{6} g_{X} \,{\widetilde g}^2, \\
{ (4\pi)^2 \beta_{\widetilde{ g}} }&=& \frac{41}{6} \widetilde{g}(\widetilde{g}^2 + 2 g_Y^2) + \left( \frac{32}{3} - \frac{41}{3} x \right) g_X ( \widetilde{g}^2 + g_Y^2 ) \nonumber\\
&&+ \left( 12  - \frac{32}{3} x + \frac{41}{6} x^2 \right) g_X^2 \widetilde{g} .
\eea

Next, the one-loop beta functions for top Yukawa coupling and the Yukawa couplings of the right-handed neutrinos are
\bea
(4\pi)^2 \beta_{y_t} &=& y_t \bigg[\frac{9}{2}y^2_t -8 g^2_S -\frac{9}{4}g^2
-\frac{17}{12}g^2_Y-\Big(\frac{2}{3}
-\frac{5}{3}x +\frac{17}{12}x^2\Big) g_X^2  \nonumber\\
&&-\Big(\frac{5}{3}-\frac{17}{6}x\Big)g_X{\widetilde g}-\frac{17}{12}{\widetilde g}^2\bigg], \\
(4\pi)^2\beta_{y_{N_i}}&=& y_{N_i} \Big(4 y^2_{N,i}+2 {\rm Tr}(y^2_N)-6 g^2_X\Big). 
\eea

Finally, the one-loop beta functions of the parameters in the potential are
\bea
(4\pi)^2 \gamma_{m^2_H}&=& m^2_H\Big(12\lambda_H+6 y^2_t -\frac{9}{2}g^2-\frac{3}{2}g^2_Y -\frac{3}{2}\big({\widetilde g}-g_X x\big)^2\Big)+2\lambda_{H\Phi} m^2_\Phi, \\
(4\pi)^2\gamma_{m^2_\Phi}&=& m^2_\Phi\Big(8\lambda_\Phi + 4{\rm Tr}(y^2_N)-24 g^2_{X}\Big) +4 \lambda_{H\Phi} m^2_H, 
\eea
and
\bea
(4\pi)^2\beta_{\lambda_H}&=& 24\lambda^2_H-6 y^4_t+\frac{9}{8}g^4+\frac{3}{8}g^4_Y+\frac{3}{4}g^2 g^2_Y +\frac{3}{4} ({\widetilde g}-g_X x )^2 ( g^2 +  g^2_Y) \nonumber\\
&& +\frac{3}{8}({\widetilde g}-g_X x)^4
+\lambda_H\Big(12 y^2_t -9g^2-3g^2_Y -3\big({\widetilde g}-g_X x\big)^2\Big) + \lambda^2_{H\Phi}, \\
(4\pi)^2\beta_{\lambda_\Phi}&=& 20\lambda^2_\Phi -16{\rm Tr}(y^4_N)+96 g^4_{X} + 8\lambda_\Phi {\rm Tr}(y^2_N)-48 \lambda_\Phi g^2_X+2\lambda^2_{H\Phi}, \\
(4\pi)^2\beta_{\lambda_{H\Phi}}&=& \lambda_{H\Phi}\Big(12\lambda_H + 8\lambda_\Phi+4\lambda_{H\Phi}+6 y^2_t -\frac{9}{2} g^2-\frac{3}{2} g^2_Y
-\frac{3}{2}({\widetilde g}-g_X x)^2 \nonumber \\
&&+ 4{\rm Tr}(y^2_N)-24 g^2_{X}\Big) +12 \big({\widetilde g}-g_X x\big)^2 g^2_{X}.
\eea


\begin{thebibliography}{999}

\bibitem{HiggsLHC}
  G.~Aad {\it et al.}  [ATLAS Collaboration],
  Phys.\ Lett.\ B {\bf 716} (2012) 1  [arXiv:1207.7214 [hep-ex]].
%
  S.~Chatrchyan {\it et al.}  [CMS Collaboration],
  Phys.\ Lett.\ B {\bf 716} (2012) 30  [arXiv:1207.7235 [hep-ex]].

\bibitem{degrassi12}
  G.~Degrassi, S.~Di Vita, J.~Elias-Miro, J.~R.~Espinosa, G.~F.~Giudice, G.~Isidori and A.~Strumia,
  JHEP {\bf 1208} (2012) 098  [arXiv:1205.6497 [hep-ph]].

\bibitem{3loops}
  K.~G.~Chetyrkin and M.~F.~Zoller,
  JHEP {\bf 1206} (2012) 033
  [arXiv:1205.2892 [hep-ph]];
  JHEP {\bf 1304} (2013) 091
  [arXiv:1303.2890 [hep-ph]];
  A.~V.~Bednyakov, A.~F.~Pikelner and V.~N.~Velizhanin,
  arXiv:1303.4364 [hep-ph].



\bibitem{sher88}
  M.~Sher,
  Phys.\ Rept.\  {\bf 179} (1989) 273.

\bibitem{shapos09}
  M.~Shaposhnikov and C.~Wetterich,
  Phys.\ Lett.\ B {\bf 683} (2010) 196  [arXiv:0912.0208 [hep-th]].
  M.~Holthausen, K.~S.~Lim and M.~Lindner,
  JHEP {\bf 1202} (2012) 037  [arXiv:1112.2415 [hep-ph]].




\bibitem{CW}
  S.~R.~Coleman and E.~J.~Weinberg,
  Phys.\ Rev.\ D {\bf 7} (1973) 1888.

\bibitem{hemp96}
  R.~Hempfling,
  Phys.\ Lett.\ B {\bf 379} (1996) 153  [hep-ph/9604278].


\bibitem{nicolai}
  K.~A.~Meissner and H.~Nicolai,
  Phys.\ Lett.\ B {\bf 648} (2007) 312
  [hep-th/0612165].


\bibitem{wu}
  W.~-F.~Chang, J.~N.~Ng and J.~M.~S.~Wu,
  Phys.\ Rev.\ D {\bf 75} (2007) 115016
  [hep-ph/0701254 [HEP-PH]].


\bibitem{iso09}
  S.~Iso, N.~Okada and Y.~Orikasa,
  Phys.\ Lett.\ B {\bf 676} (2009) 81
  [arXiv:0902.4050 [hep-ph]];
  S.~Iso, N.~Okada and Y.~Orikasa,
  Phys.\ Rev.\ D {\bf 80} (2009) 115007
  [arXiv:0909.0128 [hep-ph]];

\bibitem{iso12}
  S.~Iso and Y.~Orikasa,
  PTEP {\bf 2013} (2013) 023B08
  [arXiv:1210.2848 [hep-ph]].

\bibitem{englert13}
  C.~Englert, J.~Jaeckel, V.~V.~Khoze and M.~Spannowsky,
  arXiv:1301.4224 [hep-ph].


\bibitem{B-L}
  L.~Basso, S.~Moretti and G.~M.~Pruna,
  Phys.\ Rev.\ D {\bf 82} (2010) 055018
  [arXiv:1004.3039 [hep-ph]].



\bibitem{Sthreshold}
  J.~Elias-Miro, J.~R.~Espinosa, G.~F.~Giudice, H.~M.~Lee and A.~Strumia,
  JHEP {\bf 1206} (2012) 031
  [arXiv:1203.0237 [hep-ph]].


\bibitem{DI}
  S.~Davidson and A.~Ibarra,
  Phys.\ Lett.\ B {\bf 535} (2002) 25
  [hep-ph/0202239].



\bibitem{finetune}
  M.~Farina, D.~Pappadopulo and A.~Strumia,
  arXiv:1303.7244 [hep-ph].



\bibitem{lowerMN}
  G.~F.~Giudice, A.~Notari, M.~Raidal, A.~Riotto and A.~Strumia,
  Nucl.\ Phys.\ B {\bf 685} (2004) 89
  [hep-ph/0310123].

\bibitem{hashimoto13}
  M.~Hashimoto, S.~Iso and Y.~Orikasa,
  arXiv:1310.4304 [hep-ph].


\end{thebibliography}
\end{document}